\documentclass[prx,aps, nobalancelastpage,superscriptaddress,nolongbibliography, onecolumn]{revtex4-2}

\usepackage{color,amsthm,amsmath,amsxtra,amsfonts,dsfont,graphicx,bm,amssymb}
\usepackage[colorlinks=true,linkcolor=blue, citecolor=blue, urlcolor=blue, bookmarks]{hyperref}
\usepackage{centernot}
\usepackage[dvipsnames]{xcolor}
\usepackage{graphicx}
\usepackage{tikz}
\usepackage{braket}
\usepackage{multirow}
\usepackage{comment}
\usepackage{makecell}

\newcommand{\be}{\begin{equation}}
	\newcommand{\ee}{\end{equation}}
\newcommand{\bea}{\begin{eqnarray}}
	\newcommand{\eea}{\end{eqnarray}} 
\newcommand{\bse}{\begin{subequations}}
	\newcommand{\ese}{\end{subequations}}

\theoremstyle{plain}

\newcommand{\tr}{\mathrm{Tr}}

\usepackage{pifont}% http://ctan.org/pkg/pifont
\usepackage{tikz}
\usetikzlibrary{decorations.pathreplacing,calligraphy,decorations.markings}

\begin{document}
	\date{\today}

	\newcommand{\bbra}[1]{\<\< #1 \right|\right.}
	\newcommand{\kket}[1]{\left.\left| #1 \>\>}
	\newcommand{\bbrakket}[1]{\< \Braket{#1} \>}
	\newcommand{\pll}{\parallel}
	\newcommand{\nn}{\nonumber}
	\newcommand{\transp}{\text{transp.}}
	\newcommand{\nor}{z_{J,H}}
	
	\newcommand{\hL}{\hat{L}}
	\newcommand{\hR}{\hat{R}}
	\newcommand{\hQ}{\hat{Q}}

    \newcommand{\SD}[1]{{\color{purple} #1}}

\title{Correlation and entanglement dynamics of free fermions in disguise
}

\begin{abstract}
We study the nonequilibrium dynamics following a quantum quench in spin chains that can be solved via a mapping to free fermions in disguise. These models feature an exponential degeneracy of all energy eigenvalues, raising the question of the validity of the established framework describing the properties of integrable systems out of equilibrium. We present two main results. First, we develop an analytic method to compute the quasi-momentum distribution function characterizing the generalized Gibbs ensemble, and derive an analytic formula to compute the corresponding expectation values for special observables. Second, we adapt the standard formula for the entanglement growth based on the quasi-particle picture, discussing how our constructions do not explicitly make use of the zero-energy auxiliary free fermions responsible of the exponential degeneracies. We test our theoretical predictions against numerical tensor-network computations for different initial states and Hamiltonian parameters. For the local observables, we find excellent agreement. For the entanglement dynamics, we find small deviations suggesting that the standard quasi-particle picture is only approximately correct for the initial state considered. Our results represent a first step towards the extension of the established framework of integrable systems out of equilibrium to models hosting free fermions in disguise. 
\end{abstract}

\author{D\'avid Sz\'asz-Schagrin}
\affiliation{Dipartimento di Fisica e Astronomia, Universit\`a di Bologna and INFN, Sezione di Bologna, via Irnerio 46, 40126 Bologna, Italy}

\author{Pablo Bayona-Pena}
\affiliation{Dipartimento di Fisica e Astronomia, Universit\`a di Bologna and INFN, Sezione di Bologna, via Irnerio 46, 40126 Bologna, Italy}

\author{Lorenzo Piroli}
\affiliation{Dipartimento di Fisica e Astronomia, Universit\`a di Bologna and INFN, Sezione di Bologna, via Irnerio 46, 40126 Bologna, Italy}

\author{Eric Vernier}
\affiliation{Laboratoire de Probabilités, Statistique et Modélisation \& CNRS, Université Paris Cité, Sorbonne Université Paris, France}

\maketitle
	
%\tableofcontents

\section{Introduction}

The past two decades have witnessed the emergence of an established framework to predict the exceptional dynamical features of integrable many-body quantum systems~\cite{calabrese2016introduction,alba2021generalized,essler2016quench, caux2016quench,vidmar2016generalized,d2016quantum}. According to this framework,  when an integrable system is brought out of equilibrium, its local properties do not relax to thermal values, but define a non-trivial non-equilibrium stationary state, which is associated with a Generalized Gibbs Ensemble (GGE)~\cite{rigol2007relaxation,vidmar2016generalized}. Equivalently, the late-time properties can be described in terms of a microcanonical ensemble corresponding to a \emph{representative eigenstate}, defined in terms of the quasi-momentum distribution function of the stable quasi-particles of the model~\cite{caux213time, bertini2014quantum, caux2016quench}.

The interest in GGEs is that they often display features that are qualitatively different from thermal ones~\cite{denardis2014solution,wouters2014quenching,pozsgay2014correlations,mestyan2015quenching,piroli2016multiparticle,piroli2016exact,bertini2016quantum,piroli2016quantum,piroli2019integrable,piroli2019integrable_II}, thus representing unusual non-equilibrium phases of matter that are experimentally observable~\cite{kinoshita2006quantum,langen2015experimental}. In addition, the representation of local stationary states in terms of quasi-particle quasi-momentum distribution functions served as the basis for the development of the generalized hydrodynamics (GHD)~\cite{bertini2016transport,castro2016emergent}. This is a powerful theory to describe integrable systems in the regime of large space-time scales, providing quantitative predictions for a variety of experimental settings~\cite{malvania2021generalized,schemmer2019generalized,moller2021extension,bouchoule2022generalized,dubois2024probing,dubois2026experimental}.

The validity of the GGE has been extensively tested in Yang-Baxter integrable models~\cite{essler2016quench,vidmar2016generalized}, and it was first established for free fermions or spin chains that can be mapped to them via the Jordan-Wigner transformation~\cite{rigol2007relaxation,calabrese2011quantum,fagotti2013reduced}. In this case, the GGE predictions could be tested against analytic computations of the real-time evolution~\cite{calabrese2012quantum,calabrese2012quantum_II,essler2016quench}, while the simplified structure of free-fermionic models made it possible to directly relate the GGE quasi-particle structure to real-time dynamical features. Notably, it has been an early discovery that the GGE quasi-particle content uniquely determines the time evolution of the entanglement entropy of subsystems in the regime of large space-time scales~\cite{fagotti2008evolution}, according to the established quasi-particle picture~\cite{calabrese2006time,calabrese2016quantum,alba2017entanglement}. These findings were later extended to interacting systems, where the GGE predictions could be tested against real-time calculations, mostly based on numerical exact-diagonalization and tensor-network techniques~\cite{fagotti2014relaxation,ilievski2015complete,piroli2016exact,piroli2017correlation,pozsgay2017excited,modak2019correlation}.

In the past few years, new types of solvable models were discovered, that do not fit the traditional framework of quantum integrability. In this work, we focus in particular on a model introduced by Fendley~\cite{fendley2019free} that, despite featuring a free-fermionic spectrum, cannot be solved by means of any JW transformation~\cite{elman2021free}, see also Refs.~\cite{fendley2007cooper,de2016integrable,feher2019curious} for earlier work in this direction. This model and later generalizations~\cite{alcaraz2020free,alcaraz2020integrable,chapman2023unified, fendley2024free,fukai2025quantum,fukai2025free,sinha2026hidden,fukai2026solving_II} are now identified with the name of free
fermions in disguise (FFD) and are fundamentally different from previously-known free-fermionic systems, raising interesting questions pertaining to the interplay between solvability and classical simulability~\cite{vona2025exact,szasz2026construction,fukai2026solving}. 

Each eigenvalue in the spectrum of FFD models is exponentially degenerate in the system size~\cite{fendley2019free,vernier2025hilbert} (see however the very recent Ref.~\cite{pozsgay2026free}). Since such degeneracy is not restricted to the ground-space, one may expect it to play a role in genuine non-equilibrium protocols. It is therefore interesting and important to ask if the established framework valid for traditional Yang-Baxter integrable systems is also adequate to capture the non-equilibrium features of FFD models. This is the question that we address in this work.

It is important to stress that, contrary to the JW counterpart, establishing the validity of the GGE in FFD models is highly non-trivial. This is true even in a simple setting such that of a quantum quench~\cite{calabrese2006time}, where the system is initialized in a translation-invariant low-entangled state, and left to evolve unitarily according to the system Hamiltonian. Indeed, the FFD mapping is non-local and non-linear, making it difficult to apply standard techniques of Gaussian states~\cite{bravyi2004lagrangian,surace2022fermionic} or even compute elementary building blocks such as the overlap of the initial state with the Hamiltonian eigenstates~\cite{piroli2014recursive,brockmann2014neel,brockmann2014overlaps,foda2016overlaps,piroli2017integrable,gombor2020boundary,gombor2021factorized,gombor2022exact}. In fact, previous work has already pointed out the difficulty to compute the real-time evolution of local observables in FFD models, with explicit results available only for a very small set of observables localized around the edge of an open chain~\cite{vona2025exact,szasz2026construction}.

Despite these difficulties, in this work we provide convincing evidence for the validity of the GGE and in its ability to predict the late-time properties of FFD models after a quantum quench. We present two main results. First, we develop an analytic method to compute the GGE quasi-momentum distribution function and derive an analytic formula to compute the corresponding expectation values for special observables, the local Hamiltonian densities. Second, we adapt the standard formula for the entanglement growth based on the quasi-particle picture. We test our theoretical predictions against numerical tensor-network computations for different initial states and Hamiltonian parameters. For the local observables, we find excellent agreement. For the entanglement dynamics, we find small deviations that we cannot unambiguously attribute to finite-time effects due to the limited accessible time scales, and that suggest our conjecture is only approximately correct. Our results represent a first step towards the extension of the established framework of integrable systems out of equilibrium to models hosting free fermions in disguise. 

The rest of this work is organized as follows. We begin in Sec.~\ref{sec:the_setting} where we introduce the FFD model and the quench protocol. Our analytic results characterizing the GGE and the corresponding correlation functions are presented in Sec.~\ref{sec:gge_predictions}, while in Sec.~\ref{sec:entanglement} we discuss how the standard formula for the entanglement growth needs to be modified for FFD models. Our theoretical predictions are tested in Sec.~\ref{sec:numerics} against numerical tensor-network calculations, while our conclusions, together with an outlook on future work, are reported in Sec.~\ref{sec:outlook}.

\section{The setting}
\label{sec:the_setting}

\subsection{The model}
We consider the following local Hamiltonian,
\be
H= \sum_{m=1}^M b_m h_m,
\label{eq:H}
\ee
defined on a one-dimensional chain of $M$ spin 1/2 particles (or qubits). Here, $b_m \in  \mathbb{R}$ are arbitrary real numbers,
while each $h_m$ is an operator supported on a neighborhood of qubit $m$. The operators $h_m$ satisfy the algebra
\begin{align}
    h_m^2 &=1, \nonumber \\[3pt]
    \{ h_m, h_n\} &=0 \quad |m-n|\leq2 , \nonumber\\[3pt]
    [ h_m, h_n] &=0 \quad |m-n| >2 . 
    \label{eq:FFD_algebra}
\end{align}
Depending on the representation of the algebra~\eqref{eq:FFD_algebra} different spin models can be obtained. In the following, we consider the representation on a total Hilbert space $\mathcal{H}_{\rm FFD}= (\mathbb{C}^2)^{\otimes M}$ given by 
\begin{equation}
    h_m = Z_{m-2}Z_{m-1}X_m, \quad m \in 1, \cdots M,
\end{equation}
where $X_m$, $Z_m$ are Pauli matrices acting on qubit $m$, with the convention $Z_{-1}= Z_{0}=\mathds{1}$.

Ref.~\cite{fendley2019free} showed that the model can be diagonalized in terms of non-local fermionic modes by introducing the transfer matrix $T_M(u)$, defined through the recursion relation 
\be
T_M(u)= T_{M-1}(u) - u b_M h_M T_{M-3}, \qquad T_M(u) T_M(-u) = P_M(u) \mathds{1},
\label{eq:TransferM}
\ee

for $m>0$, and $T_m(u)=\mathds{1}$ for $m \leq 0$. Here, $P_M(u)$ is a polynomial satisfying 
\be 
P_M(u) = P_{M-1}(u) - u^2 b_M^2 P_{M-3}(u)\,,
\label{eq:PolyM}
\ee
and we let  $P_{m}(u)=1$ for $m\leq 0$.
Introducing an extra operator $\chi$ satisfying the commutation relations
\begin{align}
[ \chi,  h_m]&= 0 \quad m<M\nonumber,  \\[3pt] 
\{ \chi,  h_M\}&=0,
\end{align}
the Hamiltonian can be rewritten as 
\be
 H = \sum_{k=1}^S \epsilon_k [ \Psi_k,  \Psi_{-k} ],
\ee
where $\{ \pm u_k\}_{k=1,\cdots,S}=\{ \pm 1/\epsilon_k\}_{k=1,\cdots,S}$ denote the roots of $P_M(u)$ defined in Eq.~\eqref{eq:PolyM}, with $S = \lfloor\frac{M+2}{3} \rfloor$. For the current representation, we choose $ \chi =Z_M$, while the raising and lowering operators are given by
\be
 \Psi_{\pm k} = \frac{1}{\mathcal{N}_k}  T_M(\pm u_k) \chi T_M(\mp u_k),
\label{eq:Psi_k}
\ee
where the proportionality constant reads~\cite{fendley2019free}
\begin{equation}
\mathcal{N}_k^2= -8 u_k P_{M-1}(u_k^2)P_M'(u_k^2)= 16 u_k P_{M-1}(u_k^2)\prod_{l=1,l\neq k}^S \left( 1 - \frac{u_k^2}{u_l^2}\right)
    \label{eq:NormFactor}
\end{equation}
and where $\Psi_k = \Psi_{-k}^\dagger$.
\par

While the coefficients $b_m$ can in principle be chosen arbitrarily, we will be interested in bulk properties in the thermodynamic limit. Accordingly, we will focus on a situation where the system is invariant under the shift of a finite number of sites, and choose the parametrization
\begin{equation}\label{eq:parametrization}
    b_{3j+1} = \sqrt{\alpha}, \quad b_{3j+2} = \sqrt{\beta}, \quad b_{3j+3} = \sqrt{\gamma}\,,
\end{equation}
for all $j\in \mathbb{Z}$. In this case, the dispersion relation for the energies $\epsilon_k$ can be characterized analytically in the thermodynamic limit~\cite{fendley2019free}, as we now review. The recursion relation between the polynomials $P_m(u)$ can be recast as 
\be 
\left( \begin{array}{c} P_m(u)\\ P_{m-1}(u) \\ P_{m-2}(u)\end{array}\right)
= \bm \nu_m(u)
\left( \begin{array}{c} P_{m-1}(u) \\ P_{m-2}(u)\\ P_{m-3}(u)\end{array}\right)
\,, 
\qquad 
\bm\nu_m(u) = \left( 
\begin{array}{ccc} 1 &  0 & -u^2 b_m^2  \\ 1 & 0 & 0 \\ 0 & 1 & 0 
\end{array}   \right)\,, \label{eq:nun_def}
\ee 
and the three-site periodicity of the model naturally leads us to considering the product $\bm \nu_{\rm cell}(u):=\bm\nu_3(u)\bm\nu_2(u)\bm\nu_1(u)$.
The three corresponding eigenvalues, written as $u^2\mu_0,u^2\mu_\pm$ satisfy $\mu_+\mu_- \mu_0 = -\alpha\beta\gamma$, and can be parametrized as  
\be 
\mu_\pm = B e^{\pm i p} \,, \qquad \mu_0 = - \frac{\alpha \beta \gamma}{B^2} \,,
\ee 
where $B$ and $p$ depend on $u=1/\epsilon$ through
\begin{eqnarray}
B^3&=&2\alpha \beta \gamma\cos p+(\alpha \beta + \alpha \gamma + \beta \gamma )B \,,
\label{eq:B-cubic}
\\
\epsilon^2(p)
&=&
\frac{(B^2-\alpha\beta)(B^2-\alpha\gamma)(B^2-\beta\gamma)}
{\alpha\beta\gamma\,B^2} \,.
\label{eq:fendley-dispersion}
\end{eqnarray}
Writing $P_M(u)$ as a linear combination of the three corresponding eigenvectors and imposing the quantization condition $P_M(\pm 1/\epsilon_k)=0$ shows that the $S\sim M/3$ energy levels $\epsilon_k \sim u_k$, are described in the thermodynamic limit by a uniform density in terms of the parameter $p\in [0,\pi]$~\cite{fendley2019free}. That is, the splitting between successive values of $k$ is proportional to $ \pi/S$ and vanishes in the thermodynamic limit.

\subsection{The quench protocol}

Following the traditional framework of quantum quenches~\cite{essler2016quench}, we consider initializing the system in a low-entangled translation-invariant state. For simplicity, we focus on the product states
\be
\label{eq:init-prod-state}
    \ket{\varphi_0} = (\cos\theta\ket{1} + \sin\theta\ket{0})^{\otimes M}\,.
\ee
Next, we consider evolving the system under the Hamiltonian~\eqref{eq:H} and set  $\ket{\varphi_t}=e^{-iHt}\ket{\varphi_0}$. We will study the time evolution of local observables $O$, focusing on the late-time regime. Note that, while the model is defined with open boundary conditions, we will focus on bulk observables at a macroscopic distance from the boundaries, so that the boundary conditions will not play a role. Crucially, we will be interested in the following order of limits
\begin{equation}\label{eq:correct_order_of_limits}
    \braket{O}_{\infty}=\lim_{t\to\infty}\lim_{M\to\infty}\braket{\varphi_t|O|\varphi_t}\,.
\end{equation}
Physically, Eq.~\eqref{eq:correct_order_of_limits} corresponds to exploring the regime where the system size is much larger than any possible time scale, which rules out revivals and recurrence effects. 

Within the established framework of integrable systems out of equilibrium, for any local observable $O$ the asymptotic values~\eqref{eq:correct_order_of_limits} are captured by a GGE, namely
\begin{equation}\label{eq:gge_predictions_FFD}
        \braket{O}_{\infty}=\lim_{M\to\infty}\tr[  \varrho_{\rm GGE} O]\,.
\end{equation}
For a free fermionic system associated with fermionic modes $\{{\Psi}_j\}$, the GGE is most conveniently expressed in terms of the occupation-number operators 
\begin{eqnarray}
    N_k=\Psi_k\Psi_{-k}\,,
\end{eqnarray}
as
\be 
  \varrho_{\rm GGE} = \frac{1}{\mathcal{Z}_{\rm GGE}} e^{-\sum_{k}\beta_k  N_k} \,, \label{eq:GGE}
\ee 
where we recall that $\Psi_k^\dagger=\Psi_{-k}$. Following Ref.~\cite{fendley2019free}, it will be convenient to work with the modified operator
\begin{equation}\label{eq:occupation_number}
    \tilde N_k=[\Psi_k,\Psi_{-k}]=2N_k-1\,,
\end{equation}
and we rewrite the GGE as
\be 
  \varrho_{\rm GGE} = \frac{1}{\mathcal{Z}_{\rm GGE}} e^{-\sum_{k}\tilde \beta_k \tilde N_k} \,,\label{eq:tilted_GGE}
\ee 
where $\tilde \beta=\beta/2$. Note that, contrary to $N_k$, the operator $\tilde{N}_k$ is not positive. Importantly, the Lagrange multipliers $\tilde \beta_k$ (and hence $ \beta_k$) are fixed by the conservation of all the quantum numbers $\tilde N_k$, namely 
\be \label{eq:lagrange}
\tilde n_k \equiv \langle \varphi_0 | \tilde  N_k |\varphi_0 \rangle = \mathrm{Tr}(\varrho_{\rm GGE}  \tilde  N_k)  = - \tanh \tilde  \beta_k \,,
\ee 
where we used $\tilde  N_k^2= \mathds{1}$. In the rest of this work, we will test the validity of the Eq.~\eqref{eq:gge_predictions_FFD} for the FFD Hamiltonian~\eqref{eq:H}.

\section{The GGE predictions}
\label{sec:gge_predictions}

As mentioned, studying the GGE in FFD models is significantly harder than in traditional models that can be mapped to free fermions via the JW transformation. There are two difficulties that we need to overcome in order to obtain the GGE predictions. First, we need to compute the Lagrange multipliers via Eq.~\eqref{eq:lagrange}. Second, we need to compute the expectation value of local observables in the right-hand-side of Eq.~\eqref{eq:gge_predictions_FFD}. 

\subsection{The GGE occupation numbers}

In order to obtain the GGE Lagrange multipliers, we need to compute the expectation values of the occupation number operators in the initial states. We are interested in 
\be 
\tilde n_k = \langle\varphi_0| \tilde N_k  |\varphi_0\rangle = \frac{\langle\varphi_0| \Psi_k \Psi_{-k} |\varphi_0\rangle - \langle\varphi_0| \Psi_{-k} \Psi_{k} |\varphi_0\rangle}{\langle\varphi_0| \Psi_k \Psi_{-k} |\varphi_0\rangle + \langle\varphi_0| \Psi_{-k} \Psi_{k} |\varphi_0\rangle}   \,.
\ee 
Note that in the last expression, the $\Psi_k$ do not need to be normalized. Alternatively,  we can write
\be 
\tilde n_k = \frac{\langle\varphi_0| \Psi(u_k) \Psi(-u_k) |\varphi_0\rangle - \langle \varphi_0| \Psi(-u_k) \Psi(u_k)|\varphi_0\rangle}{\langle\varphi_0| \Psi(u_k) \Psi(-u_k) |\varphi_0\rangle + \langle\varphi_0| \Psi(-u_k) \Psi(u_k) |\varphi_0\rangle} \equiv \frac{f_M(u_k)-f_M(-u_k)}{f_M(u_k)+f_M(-u_k)},
\label{eq:nk_unnormal}
\ee 
where we have defined $\Psi(u) = T_M(u) \chi T_M(-u)$.
As shown in Appendix~\ref{sec:fmDerivation}, using the recursion relation between the transfer matrices, and the fact that for the initial product state expectation values of the form  $ \langle \varphi_0 |h_m h_n|\varphi_0 \rangle$  vanish (for $|m-n| \leq 2$), $f_M(u)$ can be obtained recursively from the following set of equations, 
\begin{align}
f_m(u)&= u^2 b_m^2  f_{m-2}(u) + u^4 b_m^2 b_{m-1}^2 f_{m-3}(u)
\nonumber\\
&
+ P_{m-1}^2(u) - 2 u b_m P_{m-3}(u)\big(P_{m-1}(u)\langle\varphi | h_m | \varphi \rangle - \frac{b_m}{b_{m-1}} (P_{m-1}(u)-P_{m-2}(u) ) \big) \langle \varphi | h_{m-1} | \varphi \rangle,
\label{eq:f-recursion}
\end{align}
with the convention that $f_m(u)=1$, $P_m(u)=1$, $b_m=0$ for  $m\leq 0$. 

A convenient way to work through \eqref{eq:f-recursion} is to recast it as a matrix equation for the $12$-dimensional vector 
\be 
\mathbf V_m(u)= (f_m, f_{m-1},f_{m-2}, (P_m , P_{m-1}, P_{m-2})\otimes (P_m , P_{m-1} , P_{m-2})).
\label{eq:Vmdef}
\ee 
In matrix form it reads 
\be 
\mathbf V_m(u)  = \left( \begin{array}{cc} \bm\lambda_m(u) & \bm\sigma_m(u)  \\ 0 & \bm\nu_m(u) \otimes \bm\nu_m(u) \end{array}  \right) \mathbf V_{m-1}(u) \equiv \bm\Lambda_m(u) \mathbf V_{m-1}(u),
\ee 
where $\bm\lambda_m(u)$, $\bm\sigma_m(u)$ and $\bm\nu_m(u)$ are respectively matrices of size $3\times 3$, $3\times 9$ and $3\times3$, whose explicit form is given in Appendix~\ref{sec:fmDerivation}.
In particular, $\bm\nu_m(u)$ encodes the recursion relation between the polynomials $P_m(u)$ and has been defined previously, see Eq.~\eqref{eq:nun_def}.
 We note that  there is some degree of freedom in defining $\bm \sigma_m(u)$, due to the redundancies in the coefficients of the vector $\mathbf{V}_m$ (for instance, $P_{m}(u)P_{m-1}(u)$ and $P_{m-1}(u)P_{m}(u)$ are given by two different entries). We could eliminate these redundancies and instead work with a 9-dimensional vector, however we choose to work with \eqref{eq:Vmdef} to keep the structure of the tensor product.

 Because of the periodicity of the chain, we are naturally led to consider the product
\begin{equation}
\bm\Lambda_{\rm cell}(u):=\bm\Lambda_3(u)\bm\Lambda_2(u)\bm\Lambda_1(u)
:=
\begin{pmatrix}
 \bm\lambda_{\rm cell}(u)&\bm\sigma_{\rm cell}(u)\\
0&\bm\nu_{\rm cell}(u)\otimes \bm\nu_{\rm cell}(u)
\end{pmatrix}.
\label{eq:cell-matrix}
\end{equation}
As we discussed previously, the spectrum of $\bm\nu_{\rm cell}(u)$ consists in three roots $u^2 \mu_\pm, u^2 \mu_0$, with 
\be 
\mu_\pm = B e^{i\pm p} \,, \qquad \mu_0 = - \frac{\alpha \beta\gamma}{B^2} \,, \qquad |\mu_0|<B  \,.
\ee 
Similarly, the eigenvalues of $\bm\lambda_{\rm cell}(u)$ consist in pairwise products of those of $\bm\nu_{\rm cell}(u)$, namely  
\be 
u^4 \mu_+ \mu_- \,, \qquad u^4 \mu_0 \mu_- \,, \qquad u^4 \mu_+ \mu_0 \,.
\ee   
The matrix $\bm\Lambda_{\rm cell}(u)$ therefore has 5 leading eigenvalues of the form $u^4 B^2,~u^4 B^2,~u^4 B^2,u^4 B^2 e^{\pm 2ip}$, with the remaining 4 eigenvalues strictly smaller in absolute value.  
Since we would like to probe relaxation properties of local observables in the thermodynamic limit, restricting to the leading eigenvalue sector is justified, such that we can rewrite 
\be 
f_M(u) \sim u^4 B^2 M (a_{+ -}(u) + a_{-+}(u)),
\label{eq:fM_final}
\ee 
where $a_{+ -}(u), a_{-+}(u)$, defined in Appendix~\ref{sec:fmDerivation}, are fixed by the 3$\times$3 Jordan block associated with the eigenvalue $u^4 B^2$.
Inserting the asymptotic form Eq.~\eqref{eq:fM_final} in  Eq.~\eqref{eq:nk_unnormal}, we have
\be \label{eq:tilde_nk-formula}
\lim_{M\to\infty}\tilde n_k = \frac{
a_{+-}(u_k)+a_{-+}(u_k)
-
a_{+-}(-u_k)-a_{-+}(-u_k)
}{
a_{+-}(u_k)+a_{-+}(u_k)
+
a_{+-}(-u_k)+a_{-+}(-u_k)
}.
\ee 
Therefore, the occupation numbers read
\be \label{eq:nk-formula}
\lim_{M\to\infty}n_k = \frac{1}{2}\left[\frac{
a_{+-}(u_k)+a_{-+}(u_k)
-
a_{+-}(-u_k)-a_{-+}(-u_k)
}{
a_{+-}(u_k)+a_{-+}(u_k)
+
a_{+-}(-u_k)+a_{-+}(-u_k)
} +1.\right]
\ee 
Note that, in the above expressions, $k$ takes value in the interval $I=[0, \pi]$. Eq.~\eqref{eq:nk-formula} represents the main result of this section. 

\subsection{The GGE expectation values}

The computation of GGE expectation values is challenging. In fact, very little is known regarding the computation of correlation functions in FFD models~\cite{fukai2026solving,szasz2026construction}. For this reason, in this work we focus on a special family of local observables, for which we can obtain an explicit analytic formula. We consider in particular the local Hamiltonian densities
\be 
\langle  h_m \rangle_{\rm GGE}
= \frac{\mathrm{Tr}(h_m \prod_k e^{-\tilde \beta_k \tilde N_k})}{\mathrm{Tr}(\prod_k e^{-\tilde \beta_k \tilde N_k})} 
= \frac{1}{2^{M-S}}\mathrm{Tr}\left(  h_m \prod_{k=1}^S \frac{1+ \tilde n_k \tilde N_k}{2}  \right)\,.
\ee 
Note that, due to the parametrization~\eqref{eq:parametrization}, the Hamiltonian is not single-site shift invariant, and the individual Hamiltonian densities are not conserved.
Expanding the product as a polynomial, we therefore need to compute objects of the form $\frac{1}{2^M} \mathrm{Tr}(h_m \prod_{k \in \mathbb I} \tilde N_k )$, where $\mathbb I$ is any subset of $\{1\ldots S\}$. 
To compute these objects, note that $ \partial_{b_m}H=  h_m$ and take the trace over the eigenbasis $|\mathbb{S},a\rangle,\ \mathbb S= \{s_1,   s_2, \cdots, s_S \}$  with $s_i= \pm1$, and $a$ stands for the degeneracy of each eigenstate. Choosing 
 $|\mathbb I|=1$, we obtain
\begin{align}
    \frac{1}{2^M}\tr[h_m \tilde N_k]&= \frac{1}{2^M}\sum_{\mathbb S, a} \langle \mathbb S, a|\partial_{b_m}H \tilde N_k | \mathbb S, a\rangle
    = \frac{1}{2^S} \sum_{\mathbb S, q} s_k s_q \partial_{b_m} \epsilon_q= \partial_{b_m} \epsilon_k.
\end{align}

In a similar fashion, for larger subsets $\mathbb I$, the expectation values can be shown to vanish. 
Therefore, 
\be 
\langle  h_m \rangle_{\rm GGE}
= \sum_{k=1}^S  \tilde n_k ~ \partial_{b_m}\epsilon_k.
\label{eq:GGE_hm}
\ee 
The derivative with respect to the couplings $\{b_m\}$ can be related to the polynomials $P_M(u)$ using the fact that 
\be 
P_M(1/\epsilon_k)=0 ,
\label{eq:Polyzero}
\ee 
where both $P_M$ and $\epsilon_k$ depend on all the couplings $\{b_m\}$.  Indeed, taking the derivative of the above equation with respect to $b_m$ yields

\begin{equation}
    0 =-\frac{\partial_{b_m}\epsilon_k }{\epsilon^2_k}P'(1/\epsilon_k)  + (\partial P)_M(1/\epsilon_k).
\end{equation}
Inserting this expression in Eq.~\eqref{eq:GGE_hm}, we obtain
\be \label{eq:gge-pred}
\langle  h_m \rangle_{\rm GGE}
= \sum_{k=1}^S  \tilde n_k \epsilon_k^2
\frac{(\partial P)_M(1/\epsilon_k)}{P_M'(1/\epsilon_k)},
\ee 
where the polynomials $(\partial P)_M$ mean the derivative of the polynomial $P
_M(u)$ with respect to $b_m$ at fixed $u$. These polynomials are defined taking the derivative of Eq.~\eqref{eq:PolyM} with respect to $b_m$, thus yielding the following recursion relations 
\bea
(\partial P)_j &=& 0 \,,\qquad j<m \\[3pt]
(\partial P)_m &=& - 2 u^2 b_m P_{m-3}(u) \,,\\[3pt]
(\partial P)_j(u) &=& (\partial P)_{j-1}(u) - u^2 b_j^2 (\partial P)_{j-3}(u)  \,, \qquad j>m\,.
\eea 
Eq.~\eqref{eq:gge-pred} represents the main result of this section, yielding an explicit prediction for the late time dynamics of the observable $\braket{h_m}$.  We expect that higher order correlation functions of the $h_m$ can be expressed in a similar fashion in terms of higher order derivatives, but leave this question for future work.

\section{The entanglement dynamics and the quasi-particle picture}
\label{sec:entanglement}

For traditional integrable models, the GGE quasi-momentum distribution function allows us to provide predictions on the entanglement dynamics of the  system after the quench. This is done using an intuitive, semi-classical quasi-particle picture originally introduced in Ref.~\cite{calabrese2005evolution} and by now well established in a variety of integrable models~\cite{calabrese2016quantum,alba2017entanglement,alba2018entanglementdynamics,bertini2018entanglement, bastianello2018spreading,alba2019entanglement,modak2019correlation,bastianello2020entanglement,klobas2021entanglement,lagnese2022entanglement,bertini2022growth}. In this section, we show that this framework can be adapted to FFD models.

It is important to note that integrable systems are usually defined with periodic boundary conditions, where the single-particle quantum number $k$ has a direct physical meaning. For example, for a simple fermionic hopping model with periodic boundaries, the single-mode quantum number $k$ can be interpreted as an actual momentum and, in the thermodynamic limit, $k\in [-\pi, \pi]$. For open boundary conditions, the allowed values of $k$ are only positive, $k\in[0,\pi]$. However, for bulk observables, an open-chain momentum-distribution function corresponds to an even periodic-chain momentum distribution function. Therefore, even for a simple free fermionic model with open boundary conditions, one can interpret the quantum number $k$ in terms of particle momenta, after symmetrizing the distribution $n_k$ around the origin. 

\begin{figure}
    \centering
    \includegraphics[width=0.4\linewidth]{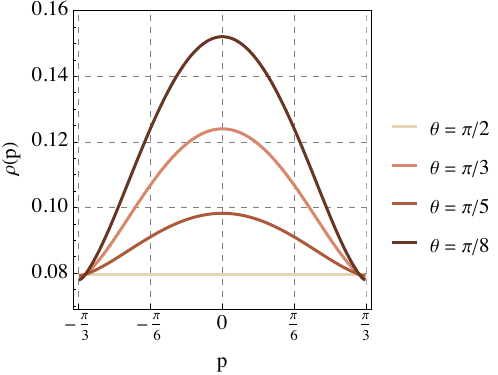}
    \caption{The quasi-particle momentum distribution $\rho(p)$ corresponding to different states \eqref{eq:init-prod-state}. For $\theta = \pi/2$, the distribution is flat with a value of $\rho(p) = \frac{1}{4 \pi}$, corresponding to the infinite-temperature state. The couplings $(\alpha, \beta, \gamma)$ are chosen to be homogeneous $(\alpha, \beta, \gamma) = (1, 1, 1)$.}
    \label{fig:rhok}
\end{figure}

For FFD models the identification of the quantum numbers $k$ with the physical momenta $p$ is less straightforward. Ref.~\cite{pozsgay2026free} studies a model that continuously extrapolates between a Jordan-Wigner solvable model and the FFD model considered here. They identify the correspondence between the physical momenta $p$ and the quantum numbers $k$ in the free fermionic point, and through continuity they argue that in the homogeneous case (where all couplings $b_m$ are set equal), the physical momentum is
\begin{eqnarray}
    p = \frac{k}{3}\,,\quad p\in\left[-\frac{\pi}{3}, \frac{\pi}{3}\right]\,,
\end{eqnarray}
corresponding to the momenta on an even periodic chain.
We therefore introduce the momentum distribution function
\begin{eqnarray}\label{eq:symmetric}
    \rho(p)=\frac{1}{2\pi}n_{|p|}\,,
\end{eqnarray}
where now $p\in [-\pi/3,\pi/3]$. Note that this distribution function is defined in such a way that
\begin{eqnarray}
    \int_{-\pi/3}^{+\pi/3}\rho(p)d{p}= D=\frac{1}{M}\sum_{k = 1}^{S}\braket{\Psi_k\Psi_{-k}}\leq \frac{S}{M}\,.
\end{eqnarray}
Accordingly, since $0\leq D\leq 1/3$, one has
\begin{eqnarray}\label{eq:density-bounds}
    0\leq \rho(p)\leq \frac{1}{2\pi}\,.
\end{eqnarray}
The function $\rho(p)$ can be interpreted as the bulk quasi-particle momentum distribution function. 
An example of the GGE quasi-particle distribution function evaluated via Eqs.~\eqref{eq:nk-formula} and~\eqref{eq:symmetric}, is reported, for different initial states, in Fig.~\ref{fig:rhok}. The function $\rho(p)$ is the starting point for the application of the quasi-particle picture for entanglement dynamics.

We recall that the entanglement between a subsystem $A$ and the rest of the system $\bar{A}$ is quantified by the entanglement entropy $S_A(t)$, defined as
\begin{equation}
S_{A}(t)=-{\rm Tr}[\rho_A(t)\log\rho_A(t)]\,,
\end{equation}
where $\rho_A(t)$ is the reduced density matrix $\rho_A(t)={\rm Tr}_{\bar A}(|\psi(t)\rangle\langle \psi(t)|)$. It represents the amount of quantum correlations between distinct spacial regions. 

Within the quasi-particle picture~\cite{calabrese2005evolution}, one models the quench as a process creating quasi-particles in correlated pairs in each point of the system. For $t > 0$, the quasi-particles move in opposite directions with
opposite velocities $\pm v(k)$, where $k$ is the quasi-particle quasi-momentum. When moving through the system, the quasi-particles correlate regions which were initially uncorrelated. The entanglement entropy $S_{A}$ is then obtained by counting the number of pairs with one quasi-particle in $A$ and the other in $\bar{A}$. Denoting by $\ell$ the number of qubits in $A$, we obtain the formula
\begin{equation}\label{eq:entropy-qp-pic}
    S_\ell(t)=t\int_{|v(p)|t\leq \ell} |v(p)|s(p)dp+\ell\int_{|v(p)|t> \ell} s(p)dp\,.
\end{equation}

The velocity $v(p)$ and the entropy $s(p)$ can be related to the quasi-particle spectrum and momentum distribution functions. In the traditional formulation, the quasi-particle velocity is obtained as
\begin{eqnarray}\label{eq:velocity}
    v(p)=\epsilon'(p)\,,
\end{eqnarray}
while $s(p)$ is given by the Yang-Yang entropy
\begin{eqnarray}\label{eq:yy-entropy-ff}
    s_{\rm YY}(p)=\rho_{\rm t} \log \rho_{\rm t}-\rho(p) \log \rho(p)-\left(\rho_{\rm t}-\rho(p)\right) \log \left(\rho_{\rm t}-\rho(p)\right)\,,
\end{eqnarray}
where $\rho_{\rm t} = \frac{1}{2 \pi}$ is the total density of the available ``slots" that can be occupied by the momenta. 

Since the velocity $v(p)$ only depends on the spectrum of the model, we expect that Eq.~\eqref{eq:velocity} continues to hold in FFD models. Note, however, that the energy of excitations is $2\epsilon_k$ since $H = \sum_k\epsilon_k[\Psi^\dagger_k,\Psi_k]=\sum_k\epsilon_k (2\Psi^\dagger_k\Psi_k - 1)$. On the other hand, since due to the bound \eqref{eq:density-bounds} on the density, we have $\rho_{\rm t} = \frac{1}{2 \pi}$, the Yang-Yang entropy formula \eqref{eq:yy-entropy-ff} is left unchanged.

In order to check the consistency of formula, we can see what happens for an infinite-temperature state, associated to a flat fermionic distribution
\begin{equation}
    \rho(k)=\frac{1}{2\cdot 2 \pi}\,,
\end{equation}
so that $D=1/6$. It is immediate to see that we get 
\begin{equation}
    s(k)=\frac{1}{2\pi}\log 2\,,
\end{equation}
which is what we should for the infinite-temperature state. In the next section, we will test the validity of the quasi-particle picture with the modified single-particle entropy~\eqref{eq:yy-entropy-ff}. 

We conclude this section discussing the role of the exponential degeneracies. In Ref.~\cite{vernier2025hilbert}. it was shown that such degeneracies can be understood in terms of auxiliary fermions with zero energy, all individually commuting with the fermions $\Psi_k$ and with the operators $h_m$. In principle, one could wonder whether these fermions should be taken into account in the construction of the GGE or in the entanglement quasi-particle picture. 

First, regarding the GGE, it is easy to convince ourselves that the auxiliary fermions do not play any role in the GGE predictions for the expectation value of $h_m$, since they all commute with it. Alternatively,  this follows from Eq.~\eqref{eq:gge-pred}: each auxiliary fermion contributes a vanishing amount, as they have zero energy. We note, however, that it could be possible that one may need to include them for the computation of more general observables. 

Regarding the entanglement dynamics, we note that auxiliary free fermions do not fit naturally within the quasi-particle picture for the entanglement dynamics. Indeed, having zero energy their velocity is vanishing, so that they do not contribute to the entanglement dynamics according to the standard quasi-particle picture. Therefore, once again, they can be neglected.

\section{Numerical results}
\label{sec:numerics}

In this section we test the predictions obtained in the previous section against numerical tensor-network computations. Here, we only present our results, while we refer to Appendix~\ref{sec:numerics_appendix} for further details on the numerical implementations. 

We first test the validity of the GGE to predict the late-time limit of local correlation functions. We focus on the dynamics of $h_m$ in the bulk of the chain. Although the system is free fermionic, we cannot compute the time evolution analytically (or simulate it efficiently) since there is no known expansion of the $h_m$ in terms of the fermionic operators $\{\Psi_{\pm k}\}$. Accordingly, we employ tensor-network simulations (namely, time-evolving block decimation, TEBD~\cite{schollwock2011density}) to probe the dynamics to finite times. In what follows, we set the system size to $M = 80$, which allows us to obtain results for the finite-time dynamics that correspond to the infinite system.

Here we present our results for $\theta = \pi/8$ and couplings $(\alpha, \beta, \gamma) = (1, 2, 3)$. Additional results corresponding to different initial states and choices of couplings are reported in Appendix~\ref{sec:additional_numerical_results_appendix} for the sake of clarity.

\begin{figure}[h]
    \centering
    \includegraphics[width=1.\linewidth]{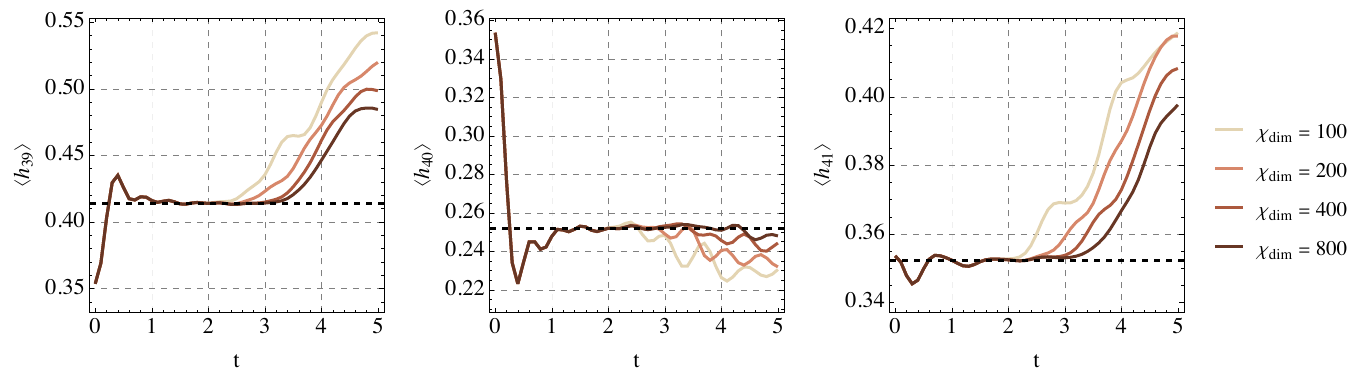}
    \caption{Time evolution of some $h_{m}$ for $M = 80$, starting from a homogeneous product state \eqref{eq:init-prod-state} with $\theta = \pi / 8$. Darker solid lines denote larger maximum bond dimensions in the TEBD. The GGE predictions (black dashed line) are given by Eq. \eqref{eq:gge-pred} together with \eqref{eq:nk-formula}, and use roots $u_k$ computed at $M = 420$. The staggered couplings are $(\alpha, \beta, \gamma) = (1, 2, 3)$. }
    \label{fig:tebd-vs-gge}
\end{figure}

Fig. \ref{fig:tebd-vs-gge} displays the real-time evolution of $h_m$ for $m = 39, 40 \text{ and } 41$ (that is, the middle of the chain) for different values of the maximum bond dimensions $\chi_{\rm dim} = 100\,\dots\, 800$ in the TEBD. Our results are reliable up to $t \approx 3$, where the curves for the two largest bond dimensions start to deviate from each other. As mentioned before, up to these times the results correspond to the thermodynamic limit, justifying the comparison with our GGE predictions, also shown in Fig. \ref{fig:tebd-vs-gge}. Clearly, the curves obtained from the numerics show a fast relaxation to a stationary value that matches with the predictions to very high precision. Together with the additional results reported in Appendix~\ref{sec:additional_numerical_results_appendix}, these findings represent strong evidence  that for the observables and initial states studied here the system relaxes to the fermionic GGE.

\begin{figure}[h]
    \centering
    \includegraphics[width=0.7\linewidth]{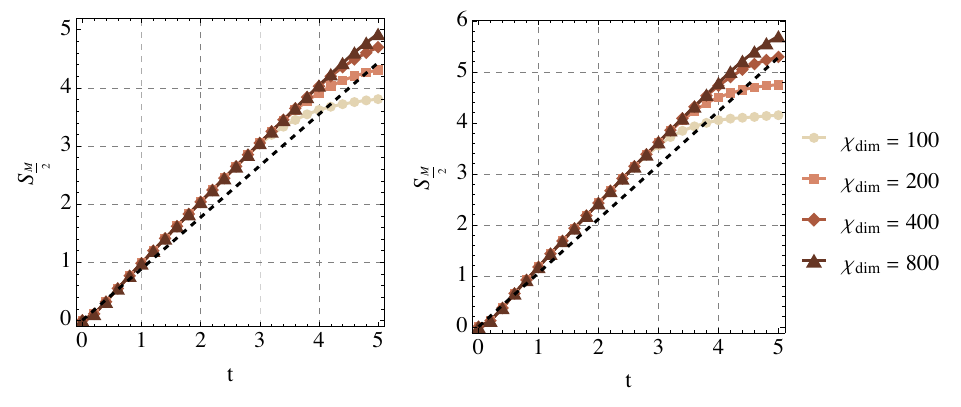}
    \caption{The dynamics of the bipartite von Neumann entanglement entropy starting from a homogeneous product state \eqref{eq:init-prod-state} with $\theta = \pi / 8$ (\textit{left}) and  $\theta = \pi / 3$ (\textit{right}) with $M = 80$. Darker solid lines denote larger maximum bond dimensions in the TEBD. The black dashed line denotes the quasi-particle picture prediction \eqref{eq:entropy-qp-pic_rate}. The staggered couplings are chosen to be $(\alpha, \beta, \gamma) = (1, 1, 1)$.}
    \label{fig:tebd-entropy}
\end{figure}

Next, we test the validity of the quasi-particle picture discussed in Sec.~\ref{sec:entanglement}. The entanglement entropy of some subsystem $A$ can be easily extracted from the TEBD simulations. We plot the time evolution of the bipartite von Neumann entropy in Fig. \ref{fig:tebd-entropy} for different values of the bond dimension. Again, we set $M = 80$ and $(\alpha, \beta, \gamma) = (1,1, 1)$, and use two different initial states with $\theta = \pi/8$ and $\theta = \pi/3$. To compare the numerics to the predictions of Sec. \ref{sec:entanglement}, we take the subsystem $A$ large, namely the first $M/2$ sites of the chain. Ideally, we would like to compare against the prediction for the asymptotic rate
\begin{equation}\label{eq:entropy-qp-pic_rate}
   \lim_{M\to\infty} (S_{M/2}(t)/t)=\int_{|v(k)|t\leq \ell} |v(k)|s(k)dk\,,
\end{equation}
for an infinite bipartite system. However, the constraints of the TEBD forbid us to probe the long-time dynamics, and we cannot make an unambiguous test of Eq.~\eqref{eq:entropy-qp-pic_rate}. Similarly, it is hard to probe the stationary value of the entanglement of finite (but large enough) intervals. For this reason, we content ourselves with a semi-quantitative comparison at the available time scales, shown in Fig.~\ref{fig:tebd-entropy}.

It is apparent from the figure that the entanglement entropy displays a linear growth  for short times and its slope is  consistent with the quasi-particle picture prediction. However, while we cannot rule out finite-time effects as Eq.~\eqref{eq:entropy-qp-pic_rate} applies only in the infinite-time limit, these plots seem to suggest some systematic deviations.  We believe that this could be due to the following reason. It is known that the quasi-particle picture captures exactly the entanglement dynamics of integrable systems only when initialized in so-called integrable states~\cite{piroli2017integrable}. These are low-entangled states made of pair of quasi-particles with opposite momenta. Conversely, corrections are expected for more general initial states~\cite{bertini2018entanglement_diagonal}. In our case, $\ket{\varphi_0}$ is not a Gaussian state and most likely not integrable (although integrability of initial states in FFD models have not been studied yet). Therefore, one would expect deviations of the quasi-particle picture, even at infinite times.

\section{Outlook}
\label{sec:outlook}

In this work we have studied the validity of the established integrability framework to describe quantum quenches in models hosting free fermions in disguise. The question is non-trivial due to the exponential degeneracy of all energy eigenvalues, potentially influencing the non-equilibrium dynamics of the system. We have developed an analytic method to compute the GGE quasi-momentum distribution function and the corresponding expectation values for special observables. In addition, we have adapted the standard formula for the entanglement growth based on the quasi-particle picture. We have tested our predictions against numerical tensor-network computations.  For the local observables, we find excellent agreement. For the entanglement dynamics, we find small deviations that we cannot unambiguously attribute to finite-time effects due to the limited accessible time scales, and that suggest our conjecture is only approximately correct.

Our results raise several questions and motivate the investigation of new directions. First, while we have tested the validity of the GGE to predict the asymptotic values of special one-point correlators, it would be interesting to study more general correlation functions. To this end, it would be important to derive a systematic approach to compute correlation functions for excited states, perhaps following general Bethe-ansatz constructions valid for interacting integrable models~\cite{pozsgay2011local, negro2013one, mestyan2014short, piroli2016multiparticle, pozsgay2017excited, bastianello2018exact}. Similarly, it would be important to test the accuracy or the limitations of our quasi-particle prediction for the entanglement dynamics more systematically.

Second, the validity of the GGE and the possibility to compute its quasi-particle probability distribution function naturally calls for the development of a GHD theory of FFD models, capturing the large-scale dynamics in inhomogeneous settings. This would complete the extension of the established integrability framework to models hosting FFDs.

Finally, we mention that our results could be extended to quantum circuits featuring FFDs. Integrable quantum circuits were shown to display GGEs with interesting non-equilibrium features that go beyond those of traditional Hamiltonian models~\cite{vanicat2018integrable,ljubotina2019ballistic,aleiner2021bethe,giudice2022temporal,claeys2022correlations,miao2023integrable,vernier2023integrable,vernier2024strong,hubner2025generalized,paletta2025integrability,paletta2025integrability_2,hubner2025generalized} and are relevant for the implementation in current digital quantum platforms~\cite{morvan2022formation,maruyoshi2023conserved,keenan2023evidence}. Therefore, it would be interesting to study the properties of GGEs in integrable circuits featuring FFDs. We leave these directions for future work.

\section*{Acknowledgments}
We are very grateful to Balázs Pozsgay for pointing out our initially unjustified identification of the single-particle quantum number $k$ with the quasi-particle momentum and for suggesting a modification of our implementation of the quasi-particle picture (as currently appearing in the draft). The work of E.V. was supported by the
ANR under grant ANR-24-CE40-7252.  This work was funded by the European Union (ERC, QUANTHEM, 101114881). Views and opinions expressed are however those of the author(s) only and do not necessarily reflect those of the European Union or the European Research Council Executive Agency. Neither the European Union nor the granting authority can be held responsible for them.

\appendix
\section{Details on analytical computations }
\label{sec:fmDerivation}
In this Appendix, we provide details on the derivation of the particle density $n_k$. The derivation is carried out by using the transfer-matrix recursion Eq.~\eqref{eq:TransferM}. We recall the definition of the one-site product state studied in this work
\begin{equation}
|\varphi_0\rangle=\bigotimes_{j=1}^M(\cos\theta\,|1\rangle+\sin\theta\,|0\rangle),
\end{equation}
for which
\begin{equation}
z:=\langle Z\rangle_{\varphi_0}=-\cos(2\theta),
\qquad
x:=\langle X\rangle_{\varphi_0}=\sin(2\theta),
\qquad
h:=\langle h_m\rangle_{\varphi_0}=z^2x=\cos^2(2\theta)\sin(2\theta),
\end{equation}
and
\begin{equation}
\langle XZ\rangle_{\varphi_0}=\langle ZX\rangle_{\varphi_0}=0.
\end{equation}

We start by extending the definition of the (unnormalized) fermion raising and lowering operators to arbitrary values of the parameter $u$,
\be 
\Psi(u) \equiv T_M(u) \chi T_{M}(u) \,.
\ee
For what follows, it is convenient to introduce the two families of  operators
\begin{align}
    A_m(u) &= T_m(u), \\
    B_m(u) &=- u b_{m+1} h_{m+1} T_{m-2}(u),
\end{align}
which obey the recursion relations 
\begin{equation}
    A_{m+1}(u)= A_m(u)+ B_m(u),  \qquad 
    B_{m+1}(u) = - u b_{m+2} A_{m-1} h_{m+2} \,,
    \label{eq:rec1}
\end{equation}
and satisfy the algebra
 \begin{align}
   [A_m(u), A_m(v)]=[B_m(u), B_m(v)]=0,\ \qquad  A_m(u)B_m(-u) + B_m(u) A_m(-u) =0  \label{eq:algebraCirq}.
 \end{align} 
As described in \cite{szasz2026construction}, the algebra \eqref{eq:algebraCirq} can be proved recursively in $m$. In particular, the second relation of the algebra can be easily proved by noting that
\begin{equation}
    A_m(u) B_m(-u)  = -P_m(u) + P_{m+1}(u)  -  B_m(u) A_m(-u)  - B_m(u) B_m(-u). 
\end{equation}

In the following, we will use the shorthand notations $A_m^\pm= A_m(\pm u)$, $B_m^\pm = B_m(\pm u)$, leaving the parameter $u$ implicit. From there, we can express the raising operator as
\begin{align}
    \Psi(u) &= A_M^+ \chi A_M^- 
      =\left( -P_M(u) +2P_{M-1}(u) + 2B_{M-1}^+ A_{M-1}^- \right)\chi.
      \label{eq:psip}
\end{align}
Likewise, for the lowering operator we obtain
\begin{align}
    \Psi(-u) &= A_M^- \chi A_M^+ = \chi\left( -P_M(u) +2P_{M-1}(u) - 2B_{M-1}^- A_{M-1}^+ \right),
      \label{eq:psim}
\end{align}
 where we used that $\{ \chi, h_M\}=0$ and $\  [ \chi, h_m]=0$ otherwise.
Combining eqs. (\ref{eq:psip}, \ref{eq:psim}) we find with a little algebra
\begin{align}
   \Psi(u)\Psi(-u) &= (P_M -2 P_{M-1})^2 -2 (P_M -2 P_{M-1}) ( B_{M-1}^+ A_{M-1}^- - B_{M-1}^- A_{M-1}^+) -4 B_{M-1}^+ A_{M-1}^-B_{M-1}^- A_{M-1}^+ .
   \label{eq:PsiPsiu}
\end{align}
We now turn to the occupation numbers, which are expressed in the main text as 
\be 
\tilde n_k = \langle\varphi_0| \tilde N_k  |\varphi_0\rangle = \frac{\langle\varphi_0| \Psi(u_k) \Psi(-u_k) |\varphi_0\rangle - \langle\varphi_0| \Psi(-u_k) \Psi(u_k) |\varphi_0\rangle}{\langle\varphi_0| \Psi(u_k) \Psi(-u_k) |\varphi_0\rangle + \langle\varphi_0| \Psi(-u_k)\Psi(u_k) |\varphi_0\rangle}   \,. 
\ee 
Inserting the expression \eqref{eq:PsiPsiu} above and using the fact that $P_M(u_k)=0$, we can therefore write
\begin{align}
  \langle {\varphi_0}|  \Psi(u_k)\Psi(-u_k)|{\varphi_0}\rangle &= 4 f_M(u_k),
\end{align}
where the functions $f_m(u)$ are defined as
\begin{equation}
    f_m(u) =  P_{m-1}^2 + P_{m-1}\langle {\varphi_0}| B_{m-1}^+ A_{m-1}^- - B_{m-1}^- A_{m-1}^+  |{\varphi_0}\rangle - \langle {\varphi_0}| B_{m-1}^+ A_{m-1}^-B_{m-1}^- A_{m-1}^+ |{\varphi_0}\rangle \,. 
    \label{eq:f_mBig}
\end{equation}
The second term of Eq.(\ref{eq:f_mBig} ) can be further simplified using the recursion relation Eq.(\ref{eq:rec1}), such that 
\begin{align}
  \langle {\varphi_0}|B_{m-1}^+ A_{m-1}^- - B_{m-1}^- A_{m-1}^+ |{\varphi_0}\rangle 
  &=- u b_m\langle {\varphi_0}| h_m A_{m-3}^+(A_{m-3}^-+ub_{m-1}h_{m-1} A_{m-4}^- +ub_{m-2}h_{m-2} A_{m-5}^-  )\nonumber \\
  &+h_m A_{m-3}^- (A_{m-3}^+ -ub_{m-1}h_{m-1} A_{m-4}^+ -ub_{m-2}h_{m-2} A_{m-5}^+  )|{\varphi_0}\rangle    \nonumber \\
  &= -2 u b_m P_{m-3}\langle{\varphi_0}| h_m|{\varphi_0}\rangle .   
\end{align}
Here we used that $\langle {\varphi_0}| h_m h_n|{\varphi_0}\rangle=0, \quad |m-n| \leq 2 $. 
Similarly, for the third therm of Eq.(\ref{eq:f_mBig} ), we find
\begin{align}
\langle{\varphi_0}|B_{m-1}^+ A_{m-1}^- B_{m-1}^- A_{m-1}^+ |{\varphi_0}\rangle &= -u^2b_m^2  \langle {\varphi_0}|h_m A_{m-3}^+A_{m-1}^- h_m A_{m-3}^-A_{m-1}^+  |{\varphi_0}\rangle.  \label{eq:BABA}
\end{align}
To further simplify the above equation, it is convenient to separate the expression in brackets into a product of two terms
\begin{align}
h_m A_{m-3}^+A_{m-1}^- &= h_m A_{m-3}^+(A_{m-3}^- +ub_{m-1}h_{m-1} A_{m-4}^- +ub_{m-2}h_{m-2} A_{m-5}^-  )  \nonumber \\
&=  P_{m-3} h_m - A_{m-3}^+ B_{m-3}^-  h_m + u b_{m-1} P_{m-4} h_m h_{m-1} + u b_{m-1} B_{m-4}^+ A_{m-4}^- h_m h_{m-1} ,
\end{align}
and
\begin{align}
h_m A_{m-3}^-A_{m-1}^+ &= h_m A_{m-3}^-(A_{m-3}^+ -ub_{m-1}h_{m-1} A_{m-4}^+ -ub_{m-2}h_{m-2} A_{m-5}^+  )  \nonumber \\
&=  P_{m-3} h_m + h_m A_{m-3}^- B_{m-3}^+   - u b_{m-1} P_{m-4} h_m h_{m-1} - u b_{m-1} B_{m-4}^- A_{m-4}^+ h_m h_{m-1} .
\end{align}
Therefore, Eq.(\ref{eq:BABA}) can be rewritten as
\begin{align}
\langle {\varphi_0}|B_{m-1}^+ A_{m-1}^- B_{m-1}^- A_{m-1}^+ |{\varphi_0}\rangle &= -u^2b_m^2 \langle \Big[ P_{m-3}^2 + P_{m-3}B_{m-3}^+  A_{m-3}^- - P_{m-3}B_{m-3}^-  A_{m-3}^+ - B_{m-3}^+  A_{m-3}^- B_{m-3}^-  A_{m-3}^+  \nonumber \\
&+ u^2 b_{m-1}^2 (P_{m-4}^2 + P_{m-4}B_{m-4}^+  A_{m-4}^- - P_{m-4}B_{m-4}^-  A_{m-4}^+   \nonumber \\&- B_{m-4}^+ A_{m-4}^- B_{m-4}^-  A_{m-4}^+ ) -2u b_{m-1} ( P_{m-3} P_{m-4} )  h_{m-1} \Big]  \rangle_{\varphi_0} \nonumber \\
&= -\left(u^2 b_m^2 f_{m-1}^2 +u^4 b_m^2 b_{m-1}^2 f_{m-3} + 2 u \frac{b_m^2}{b_{m-1}} P_{m-3}(P_{m-1} -P_{m-2}) \langle h_{m-1}\rangle_{\varphi_0}\right)
\label{eq:thirdTerm}
\end{align}
 where in the last equality, we used the recursion relation Eq.(\ref{eq:PolyM}), again $\langle {\varphi_0}| h_m h_n|{\varphi_0}\rangle=0, \quad |m-n| \leq 2 $ and the algebra of Eq.~\eqref{eq:algebraCirq}.
 Putting the above expressions together, we arrive at the recursion of the main text Eq.~\eqref{eq:f-recursion}. Next, we focus on the recursive computation of $f_M(u)$ with the convention
\begin{equation}
f_m(u)=1,\qquad P_m(u)=1,\qquad b_m=0,
\qquad m\leq 0.
\end{equation}
Introducing the three-component vectors
\begin{equation}
\mathbf f_m(u)
=
\begin{pmatrix}
f_m(u)\\
f_{m-1}(u)\\
f_{m-2}(u)
\end{pmatrix},
\qquad
\mathbf P_m(u)
=
\begin{pmatrix}
P_m(u)\\
P_{m-1}(u)\\
P_{m-2}(u)
\end{pmatrix}.
\end{equation}
Then the homogeneous part of \eqref{eq:f-recursion} is generated by
\begin{equation}
{\bm \lambda}_m(u)
=
\begin{pmatrix}
0 & u^2b_m^2 & u^4b_m^2b_{m-1}^2\\
1 & 0 & 0\\
0 & 1 & 0
\end{pmatrix},
\end{equation}
while the polynomial recursion \eqref{eq:PolyM} is generated by
\begin{equation}
{\bm  \nu}_m(u)
=
\begin{pmatrix}
1 & 0 & -u^2b_m^2\\
1 & 0 & 0\\
0 & 1 & 0
\end{pmatrix},
\qquad
\mathbf P_m(u)={\bm \nu}_m(u)\mathbf P_{m-1}(u).
\end{equation}
Since the source term in \eqref{eq:f-recursion} is quadratic in the polynomials \(P\), it is natural to introduce
\begin{equation}
\mathbf W_m(u):=\mathbf P_m(u)\otimes \mathbf P_m(u).
\end{equation}
The full twelve-dimensional vector is
\begin{equation}
\mathbf V_m(u)
=
\begin{pmatrix}
\mathbf f_m(u) ,\
\mathbf W_m(u)
\end{pmatrix}^T
=
\begin{pmatrix}
f_m,\ f_{m-1},\ f_{m-2},\
(P_m,P_{m-1},P_{m-2})\otimes(P_m,P_{m-1},P_{m-2})
\end{pmatrix}^T.
\end{equation}
It evolves as
\begin{equation}
\mathbf V_m(u)=\bm \Lambda_m(u)\mathbf V_{m-1}(u),
\end{equation}
with block upper-triangular matrix
\begin{equation}
\bm\Lambda_m(u)
=
\begin{pmatrix}
\bm\lambda_m(u) & \bm\sigma_m(u)\\
0 & \bm\nu_m(u)\otimes \bm\nu_m(u)
\end{pmatrix}.
\label{eq:Lambda-m}
\end{equation}
The \(3\times 9\) matrix \(\bm\sigma_m(u)\) encodes the polynomial source in
\eqref{eq:f-recursion}. In the ordered basis
\begin{equation}
(P_{m-1},P_{m-2},P_{m-3})
\otimes
(P_{m-1},P_{m-2},P_{m-3}),
\label{eq:tensorP}
\end{equation}
only the first row of \(\bm\sigma_m\) is nonzero. With the normalization of
\eqref{eq:f-recursion}, this row is
\begin{equation}
\bm\sigma_m^{(1)}
=
\begin{pmatrix}
1 & 0 & 0 & 0 & 0 & 0 &
-2u b_m\langle h_m\rangle_{\varphi_0}
+
2u\frac{b_m^2}{b_{m-1}}\langle h_{m-1}\rangle_{\varphi_0}
&
-2u\frac{b_m^2}{b_{m-1}}\langle h_{m-1}\rangle_{\varphi_0}
&0
\end{pmatrix},
\label{eq:mu-row}
\end{equation}
and  the Hamiltonian couplings are given by
\begin{equation}
b_{3j+1}^2=\alpha,
\qquad
b_{3j+2}^2=\beta,
\qquad
b_{3j+3}^2=\gamma.
\end{equation}
We note that the choice of this row has some degree of freedom due to the structure of the tensor product of~\eqref{eq:tensorP}.

Next we extract the five-dimensional leading sector of the matrix $\bm\Lambda_{\rm cell}$. The polynomial-bilinear block is
\begin{equation}
 \bm\nu_{\rm cell}(u)\otimes \bm\nu_{\rm cell}(u).
\end{equation}
Since the eigenvalues of \(\bm\nu_{\rm cell}(u)\) are
\begin{equation}
\rho_\pm=u^2Be^{\pm ip},
\qquad
\rho_0=-u^2\frac{s_3}{B^2},
\end{equation}
the eigenvalues of \(\bm\nu_{\rm cell}(u)\otimes \bm\nu_{\rm cell}(u)\) are
\begin{equation}
\rho_i\rho_j,
\qquad i,j\in\{+,-,0\}.
\end{equation}
The four eigenvalues with modulus \(|u^4B^2|\) are
\begin{align}
\rho_+\rho_-=u^4B^2,\quad
\rho_-\rho_+=u^4B^2,\quad
\rho_+\rho_+=u^4B^2e^{2ip},\quad
\rho_-\rho_-=u^4B^2e^{-2ip}.
\end{align}
Together with the leading real eigenvalue \(\Lambda(u)=u^4B^2\) of the \(\bm\lambda\)-cell, these form a five-dimensional sector of the full matrix \(\ \bm\Lambda_{\rm cell}(u)\).

Let
\begin{equation}
\bm\Lambda_{\rm cell} (u)|R_\lambda\rangle=\Lambda(u)|R_\lambda\rangle,
\qquad
\langle L_\lambda(u)|\bm\Lambda_{\rm cell}(u)=\Lambda\langle L_\lambda|,
\end{equation}
and
\begin{equation}
\bm\nu_{\rm cell} (u)|R_\pm\rangle=\rho_\pm|R_\pm\rangle,
\qquad
\langle L_\pm|\bm\nu_{\rm cell}(u)=\rho_\pm\langle L_\pm|.
\end{equation}
We choose the left and right eigenvectors biorthonormally,
\begin{equation}
\langle L_a|R_b\rangle=\delta_{ab}.
\end{equation}
In the ordered basis
\begin{equation}
\mathcal B_{\rm lead}
=
\left\{
|\lambda\rangle,\,
|+\!-\rangle,\,
|-\!+\rangle,\,
|+\!+\rangle,\,
|-\!-\rangle
\right\},
\end{equation}
where
\begin{equation}
|\lambda\rangle=
\begin{pmatrix}
|R_\lambda\rangle\\0
\end{pmatrix},
\qquad
|ij\rangle=
\begin{pmatrix}
0\\ |R_i\rangle\otimes |R_j\rangle
\end{pmatrix},
\end{equation}
the leading block of \(\bm\Lambda_{\rm cell}(u)\) takes the triangular form
\begin{equation}
\bm\Lambda_{\rm cell}(u)
|_{\rm lead}
=
\Lambda(u)
\begin{pmatrix}
1 & a_{+-}(u) & a_{-+}(u) & a_{++}(u) & a_{--}(u)\\
0 & 1 & 0 & 0 & 0\\
0 & 0 & 1 & 0 & 0\\
0 & 0 & 0 & e^{2ip} & 0\\
0 & 0 & 0 & 0 & e^{-2ip}
\end{pmatrix}.
\label{eq:five-block}
\end{equation}
The coefficients in the first row come entirely from the upper-right source block
\(\bm\sigma_{\rm cell}(u)\) in \eqref{eq:cell-matrix}. Explicitly,
\begin{equation}
a_{ij}(u)
=
\frac{
\langle L_\lambda|\bm\sigma_{\rm cell}(u)|R_i\rangle\otimes|R_j\rangle
}{
\Lambda(u)
},
\qquad
i,j\in\{+-, -+, ++, --\}.
\label{eq:aij-def}
\end{equation}
Next we compute the large-\(M\) asymptotics of the matrix $\Lambda_{\rm cell}$.
Letting \(M=3S\), Define the normalized leading block
\begin{equation}
\widetilde{\bm \Lambda}_{\rm cell}(u):=\frac{1}{\Lambda(u)} 
 \bm\Lambda_{\rm cell}(u)|_{\rm lead}.
\end{equation}
From \eqref{eq:five-block}, its \(S\)-th power is
\begin{equation}
\widetilde{\bm \Lambda}_{\rm cell}(u)^S
=
\begin{pmatrix}
1
&
S a_{+-}
&
S a_{-+}
&
a_{++}\dfrac{1-e^{2iSp}}{1-e^{2ip}}
&
a_{--}\dfrac{1-e^{-2iSp}}{1-e^{-2ip}}
\\
0&1&0&0&0\\
0&0&1&0&0\\
0&0&0&e^{2iSp}&0\\
0&0&0&0&e^{-2iSp}
\end{pmatrix}.
\label{eq:five-block-power}
\end{equation}
The entries proportional to \(a_{++}\) and \(a_{--}\) are bounded oscillatory geometric sums for generic \(p\). By contrast, the entries proportional to \(a_{+-}\) and \(a_{-+}\) grow linearly with the number of cells \(S\). Therefore the leading thermodynamic contribution to \(f_M(u)\) is controlled by
\begin{equation}
A(u):=a_{+-}(u)+a_{-+}(u),
\end{equation}
and
\begin{equation}
f_M(u)\sim \Lambda(u)^S\,S\,A(u),
\qquad M=3S,
\label{eq:f-asymptotic}
\end{equation}
up to boundary-overlap factors that are common to the numerator and denominator of the occupation ratio.

Finally, since the mode occupations are extracted from
\begin{equation}
\tilde{n}_k
=
\frac{f_M(u_k)-f_M(-u_k)}
{f_M(u_k)+f_M(-u_k)},
\qquad
u_k=\frac{1}{\epsilon_k},
\end{equation}
and since the exponential scale \(\Lambda(u)^S\) is the same for \(u\) and \(-u\), the leading thermodynamic occupation is
\begin{equation}
\lim_{M\to \infty}\tilde{n}_k
=
\frac{
A(u_k)-A(-u_k)
}{
A(u_k)+A(-u_k)
}
=
\frac{
a_{+-}(u_k)+a_{-+}(u_k)
-
a_{+-}(-u_k)-a_{-+}(-u_k)
}{
a_{+-}(u_k)+a_{-+}(u_k)
+
a_{+-}(-u_k)+a_{-+}(-u_k)
}.
\label{eq:nk-final}
\end{equation}

\section{Details on the numerical computations}
\label{sec:numerics_appendix}

\subsection{Time evolution of the system}
Here we present some details of the implementation of the numerical simulations. We are computing
\begin{equation}
    \braket{O(t)} = \braket{\varphi_t|O|\varphi_t}
\end{equation}
where
\begin{equation}
    \ket{\varphi_t} = e^{-iHt}\ket{\varphi_0}\,.
\end{equation}
and $O$ is some local observable.
We are employing tensor-network methods, namely the time-evolving block decimation (TEBD). In the TEBD, the initial state $\ket{\phi_0}$ is represented as a matrix product state (MPS). To evolve it, one generally applies a form of the Suzuki-Trotter decomposition of the evolution operator:
\begin{equation}
    \ket{\varphi_t} = e^{-i h_1 dt/2}e^{-i h_2 dt/2}\dots e^{-i h_{M-1} dt/2}e^{-i h_M dt}e^{-i h_{M-1} dt/2}\dots e^{-i h_{1} dt/2}\ket{\varphi_0} + \mathcal{O}(dt^3)\,.
\end{equation}
Throughout our numerics, we take $dt = 0.1$ and check that for the simulated time window the total energy of the system is conserved up to negligible errors. Following an application of each gate $e^{-i h_m dt/2}$, one performs an SVD decomposition of the (local) state, resulting in an MPS state $\ket{\varphi_{t+dt}}$ at every timestep. Then, the numerical accuracy is constrained by the bond dimension of this MPS, corresponding to the maximum entanglement (and correlations) that can be contained in the state. Both here and in the main text we disclose results obtained with increasing values of the bond dimension and they are reliable up to times where the curves for the two largest bond dimensions agree.

\subsection{The GGE expectation values of $h_m$}

To compute the GGE expectation values
\be \label{eq-app:gge-pred}
\langle  h_m \rangle_{\rm GGE}
= \sum_{k=1}^S  \tilde{n}_k \epsilon_k^2
\frac{(\partial P)_M(1/\epsilon_k)}{P_M'(1/\epsilon_k)},
\ee 
it is necessary to evaluate the polynomials $(\partial P)_M$ defined by the following recursion relations 
\bea
(\partial P)_j &=& 0 \qquad j<m \\[3pt]
(\partial P)_m &=& - 2 u^2 b_m P_{m-3}(u) \\[3pt]
(\partial P)_j(u) &=& (\partial P)_{j-1}(u) - u^2 b_j^2 (\partial P)_{j-3}(u)  \,, \qquad j>m\,.
\eea 

Although the Gibbs ensemble $\{N_k\}$ can be analytically computed in the thermodynamic limit, Eq. \eqref{eq-app:gge-pred} is defined in a finite volume $M$. To evaluate it, we numerically compute the roots $\{u_k\}$ of the polynomial $P_M(u)$ for a large enough value of $M$ (we take $M = 420$) and insert it into the formula \eqref{eq-app:gge-pred}.

\section{Additional numerical results}
\label{sec:additional_numerical_results_appendix}

Lastly, we present some additional numerical results. As in the main text, we focus on initial product states
\begin{eqnarray}
    \ket{\varphi_0} = (\cos\theta\ket{1} + \sin\theta\ket{0})^{\otimes M}
\end{eqnarray}
parametrized by the tilting angle $\theta$. While in the main text we presented results for $\theta = \pi/8$, here we include additional results for $\theta = \pi/3$.

\begin{figure}[h!]
    \centering
    \includegraphics[width=1.\linewidth]{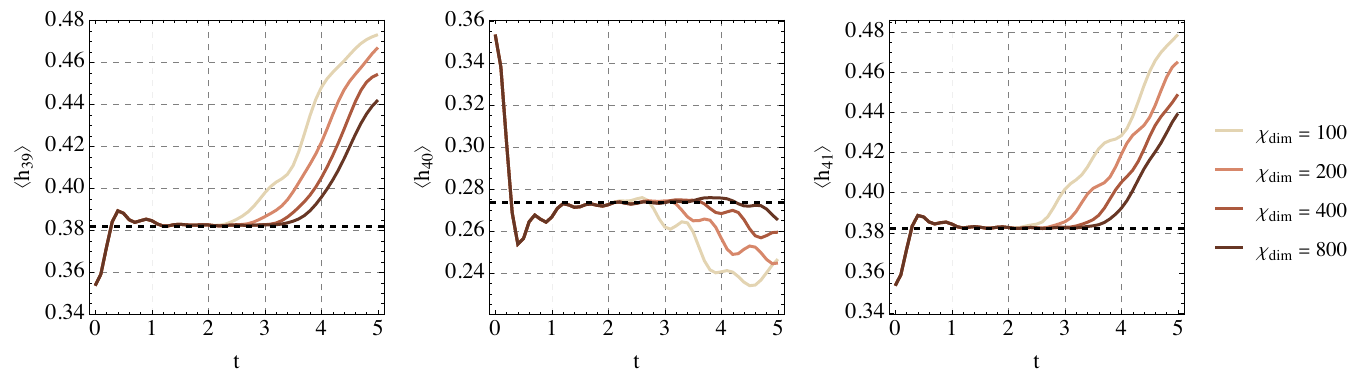}\\
    \includegraphics[width=1.\linewidth]{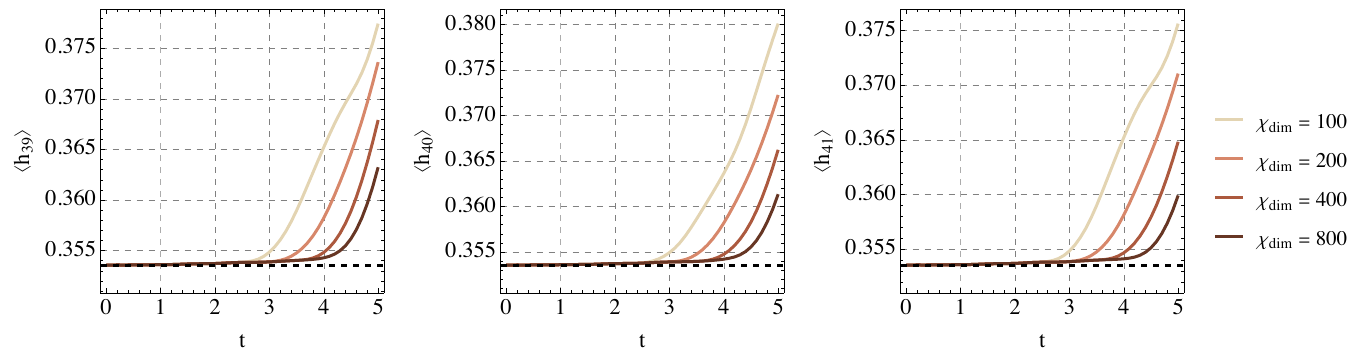}
    \caption{Time evolution of some $h_{m}$ for $M = 80$, starting from a homogeneous product state \eqref{eq:init-prod-state} with $\theta = \pi / 8$. Darker solid lines denote larger maximum bond dimensions in the TEBD. The GGE predictions (black dashed line) are given by Eq. \eqref{eq:gge-pred} together with \eqref{eq:nk-formula}, and use roots $u_k$ computed in $M = 420$. The staggered couplings (\textit{from top to bottom}) are $(\alpha, \beta, \gamma) = (1, 2, 2) \text{ and } (1,1,1)$. }
    \label{fig:tebd-vs-gge_appendix-pi8}
\end{figure}

\begin{figure}[h!]
    \centering
    \includegraphics[width=1.\linewidth]{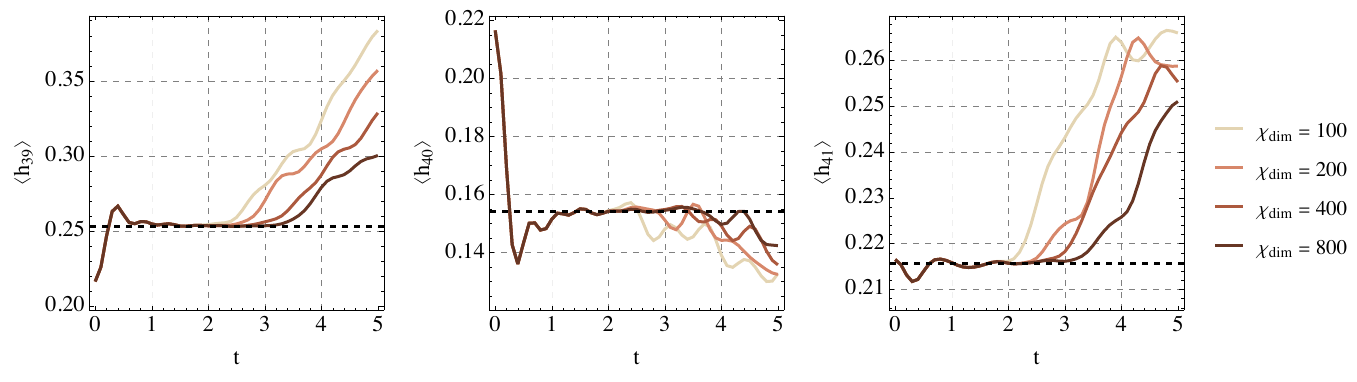}\\
    \includegraphics[width=1.\linewidth]{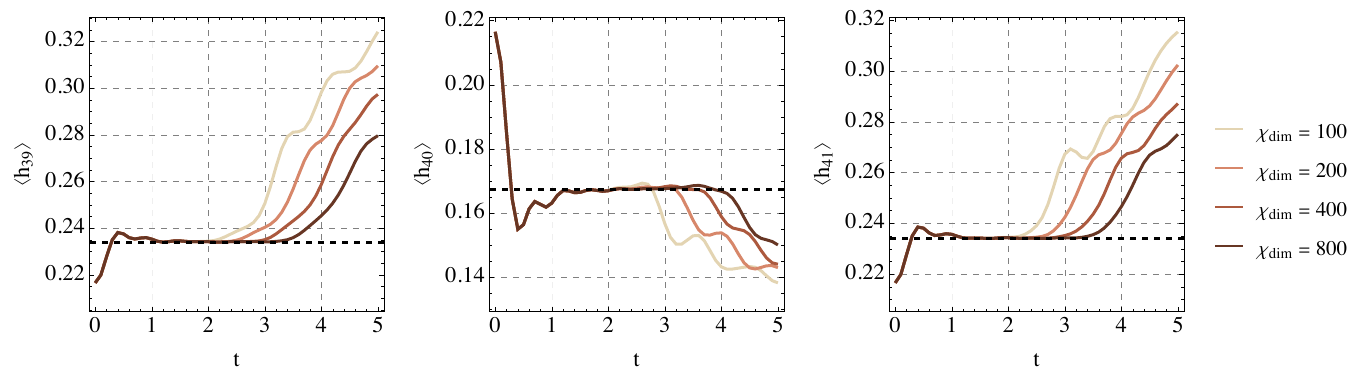}\\
    \includegraphics[width=1.\linewidth]{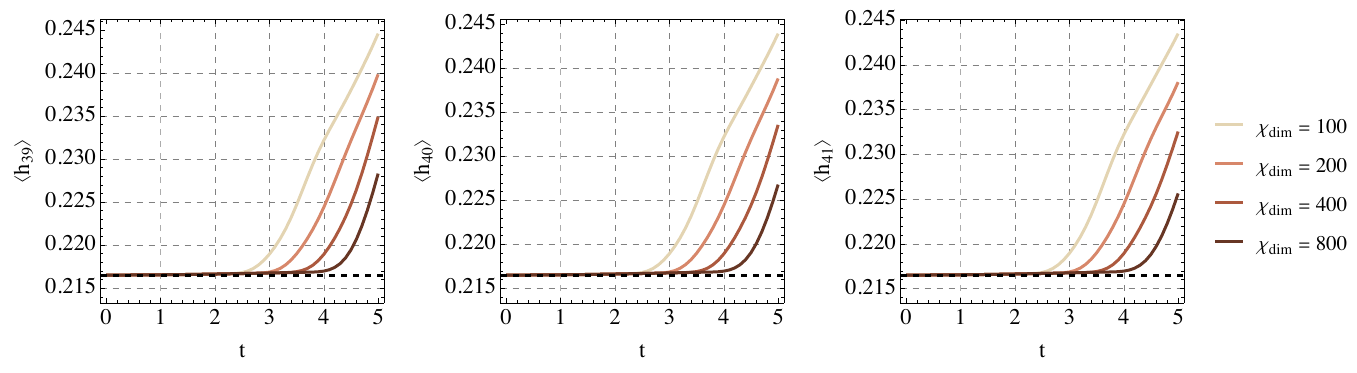}
    \caption{Time evolution of some $h_{m}$ for $M = 80$, starting from a homogeneous product state \eqref{eq:init-prod-state} with $\theta = \pi / 3$. Darker solid lines denote larger maximum bond dimensions in the TEBD. The GGE predictions (black dashed line) are given by Eq. \eqref{eq:gge-pred} together with \eqref{eq:nk-formula}, and use roots $u_k$ computed in $M = 420$. The staggered couplings (\textit{from top to bottom}) are $(\alpha, \beta, \gamma) =  (1, 2, 3),\,\,(1, 2, 2) \text{ and } (1,1,1)$. }
    \label{fig:tebd-vs-gge_appendix-pi3}
\end{figure}

For the sake of completeness, we also vary the coupling terms of the Hamiltonian. Since the couplings $b_m$ are staggered every three site
\begin{equation}
    b_{3j+1} = \sqrt{\alpha}, \quad b_{3j+2} = \sqrt{\beta}, \quad b_{3j+3} = \sqrt{\gamma}\,,
\end{equation}
we find it instructive to tune the system between the completely staggered case, where $(\alpha, \beta, \gamma)$ are different, and the homogeneous case, where every coupling is equal. Therefore, we present results for three choices of the parameters $(\alpha, \beta, \gamma)$:
\begin{eqnarray}
    (\alpha, \beta, \gamma) = \begin{cases}
       \,\, (1, 2, 3)\\
       \,\,  (1, 2, 2)\\
       \,\,  (1, 1, 1)
    \end{cases}.
\end{eqnarray}
Results for $\theta = \pi/8$ and $(\alpha, \beta, \gamma) = (1, 2, 3)$ are presented in the main text. In Fig. \ref{fig:tebd-vs-gge_appendix-pi8}, the time evolution for some $h_m$ in the middle of the chain are presented for the same state with $(\alpha, \beta, \gamma) = (1, 2, 2)$ and (1,1,1). The same is shown in Fig. \ref{fig:tebd-vs-gge_appendix-pi3} for $\theta = \pi / 3$ and the three different choices of coupling parameters. Clearly, the conclusions of the main text are also supported by these results: for these initial states and observables, the system relaxes to a fermionic GGE.

\begin{figure}[h!]
    \centering
    \includegraphics[width=0.7\linewidth]{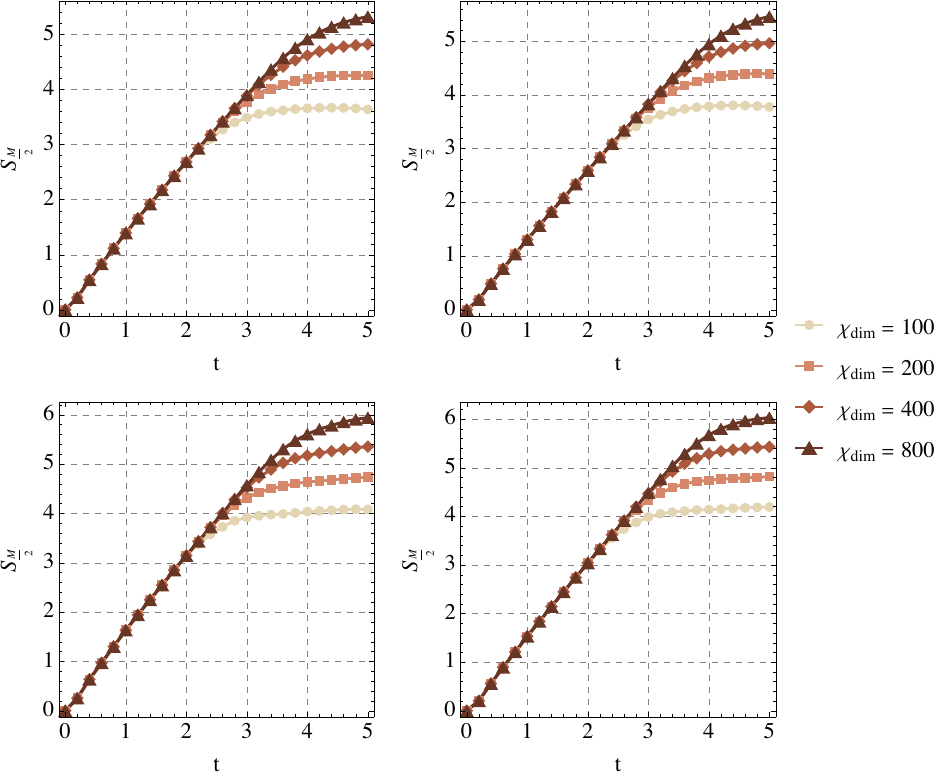}
    \caption{The dynamics of the bipartite von Neumann entanglement entropy starting from a homogeneous product state \eqref{eq:init-prod-state} with $\theta = \pi / 8$ (\textit{top}) and $\theta = \pi / 3$ (\textit{bottom}) for $M = 80$. Darker solid lines denote larger maximum bond dimensions in the TEBD. The staggered couplings are chosen to be $(\alpha, \beta, \gamma) = (1, 2, 2)$ (\textit{left}) and (1,1,1) (\textit{right}).}
    \label{fig:tebd-entropy-appendix}
\end{figure}

We also present additional results for the dynamics of the bipartite von-Neumann entropy. As before, we set $M = 80$ and look at the entanglement between the first $M/2$ sites and the rest of the system. Results for $\theta = \pi/8$ and $\theta = \pi  /3$ with $(\alpha, \beta, \gamma) = $ (1, 2, 2) and (1,1,1) are shown in Fig. \ref{fig:tebd-entropy-appendix}. The curves clearly show that for early times, the entanglement start to grow linearly with time, suggesting that some form of (modified) quasi-particle picture can describe the dynamics. However, the results presented in the main text are valid in the homogeneous system, where we can define a notion of physical momentum for the fermionic excitations. As such, we refrain ourselves from comparing our predictions with these results.
% Again, these results support the conclusions of the main text: the modified quasi-particle picture detailed in Sec. \ref{sec:entanglement} reasonably matches the early-time dynamics of the entanglement. However, due to the limitations on the available time-window in the TEBD, one can not unambiguously ascribe the small deviations to finite-time effects. 

\clearpage

\bibliography{bibliography}

\end{document}